\documentclass[conference]{IEEEtran}

\usepackage{cite}
\usepackage{amsmath}
\usepackage{caption}
\usepackage{subcaption}
\usepackage{graphicx}
\usepackage[linesnumbered,ruled,vlined]{algorithm2e}
\usepackage{url}
\usepackage{hyperref}
\usepackage{verbatim}
\usepackage{listings}
\usepackage{mathtools}
\usepackage{float}

\lstdefinelanguage{Julia}{
  basicstyle=\small\ttfamily,
  showspaces=false,
  showstringspaces=false,
  keywordstyle={\textbf},
  morekeywords={if,else,elseif,while,for,begin,end,quote,try,catch,return,local,abstract,function,stagedfunction,macro,ccall,finally,typealias,break,continue,type,global,module,using,import,export,const,let,bitstype,do,in,baremodule,importall,immutable},
  escapeinside={~}{~},
  morecomment=[l]{\#},
  commentstyle={},
  morestring=[b]",
}

\DeclarePairedDelimiter\floor{\lfloor}{\rfloor}

\begin{document}

\title{Robust benchmarking in noisy environments}

\author{\IEEEauthorblockN{Jiahao Chen and Jarrett Revels}
\IEEEauthorblockA{Computer Science and Artificial Intelligence Laboratory\\
Massachusetts Institute of Technology\\
Cambridge, Massachusetts 02139--4307\\
Email: \{jiahao,jrevels\}@csail.mit.edu}
}

\maketitle

\begin{abstract}
We propose a benchmarking strategy that is robust in the presence of timer
error, OS jitter and other environmental fluctuations, and is insensitive to
the highly nonideal statistics produced by timing measurements.
We construct a model that explains how these strongly nonideal statistics can
arise from environmental fluctuations, and also justifies our proposed
strategy. We implement this strategy in the BenchmarkTools Julia package, where
it is used in production continuous integration (CI) pipelines for developing
the Julia language and its ecosystem.
\end{abstract}

\IEEEpeerreviewmaketitle

\section{Introduction}
\label{sec:intro}

Authors of high performance applications rely on benchmark suites to detect and avoid
program regressions. However, many developers often run benchmarks and interpret their
results in an ad hoc manner with little statistical rigor. This ad hoc interpretation wastes
development time and can lead to misguided decisions that worsen performance.

In this paper, we consider the problem of designing a language- and
platform-agnostic benchmarking methodology that is suitable for continuous
integration (CI) pipelines and manual user workflows. Our methodology
especially focuses on the accommodation of benchmarks whose expected executions
times are short enough that timing measurements are vulnerable to error due to
insufficient system timer accuracy (generally on the order of microseconds or
shorter).

\subsection{Accounting for performance variations}
\label{sec:variations}

Modern hardware and operating systems introduce many confounding factors that complicate a
developer's ability to reason about variations in user space application
performance~\cite{HP5e}.\footnote{A summary of these factors can be found in the
\href{https://github.com/JuliaCI/BenchmarkTools.jl}{\lstinline|BenchmarkTools|}
documentation in the
\href{https://github.com/JuliaCI/BenchmarkTools.jl/blob/4db27210d43abf2c55226366f3a749afe1d64951/doc/linuxtips.md}{docs/linuxtips.md}
file.} Consecutive timing measurements can fluctuate, possibly in a correlated
manner, in ways which depend on a myriad of factors such as environment
temperature, workload, power availability, and network traffic, and operating
system (OS) configuration.

There is a large body of research on system quiescence aiming to identify and
control for individual sources of variation in program run time measurements,
each of which must be ameliorated in its own way. Many factors stem from OS
behavior, including CPU frequency scaling \cite{RHEL6}, address space layout
randomization (ASLR)~\cite{Shacham2004}, virtual memory
management~\cite{Oyama2014,Oyama2016}, differences between CPU privilege
levels~\cite{Zaparanuks2009}, context switches due to interrupt handling~\cite{Tsafrir2007},
activity from system daemons and cluster managers~\cite{Petrini2003}, and suboptimal
process- and thread-level scheduling~\cite{Lozi2016}. Even seemingly irrelevant
configuration parameters like the size of the OS environment can confound experimental
reproducibility by altering the alignment of data in memory~\cite{Mytkowicz2009}. Other
sources of variation come from specific language features or implementation details. For
example, linkers for many languages are free to choose the binary layout of the library or
executable arbitrarily, resulting in non-deterministic memory layouts~\cite{Georges2008}.
This problem is exacerbated in languages like C++, whose compilers introduce arbitrary name
mangling of symbols~\cite{Kalibera2005}. Overzealous compiler optimizations can also
adversely affect the accuracy of hardware counters~\cite{Zaparanuks2009}, or in extreme
cases eliminate key parts of the benchmark as dead code. Yet another example is garbage
collector performance, which is influenced from system parameters such as heap
size~\cite{Blackburn2004}.

\subsection{Statistics of timing measurements are not i.i.d.}
\label{sec:toughstats}

\begin{figure}
\centering
\begin{subfigure}{0.22\textwidth}
    \centering
    \includegraphics[width=\textwidth]{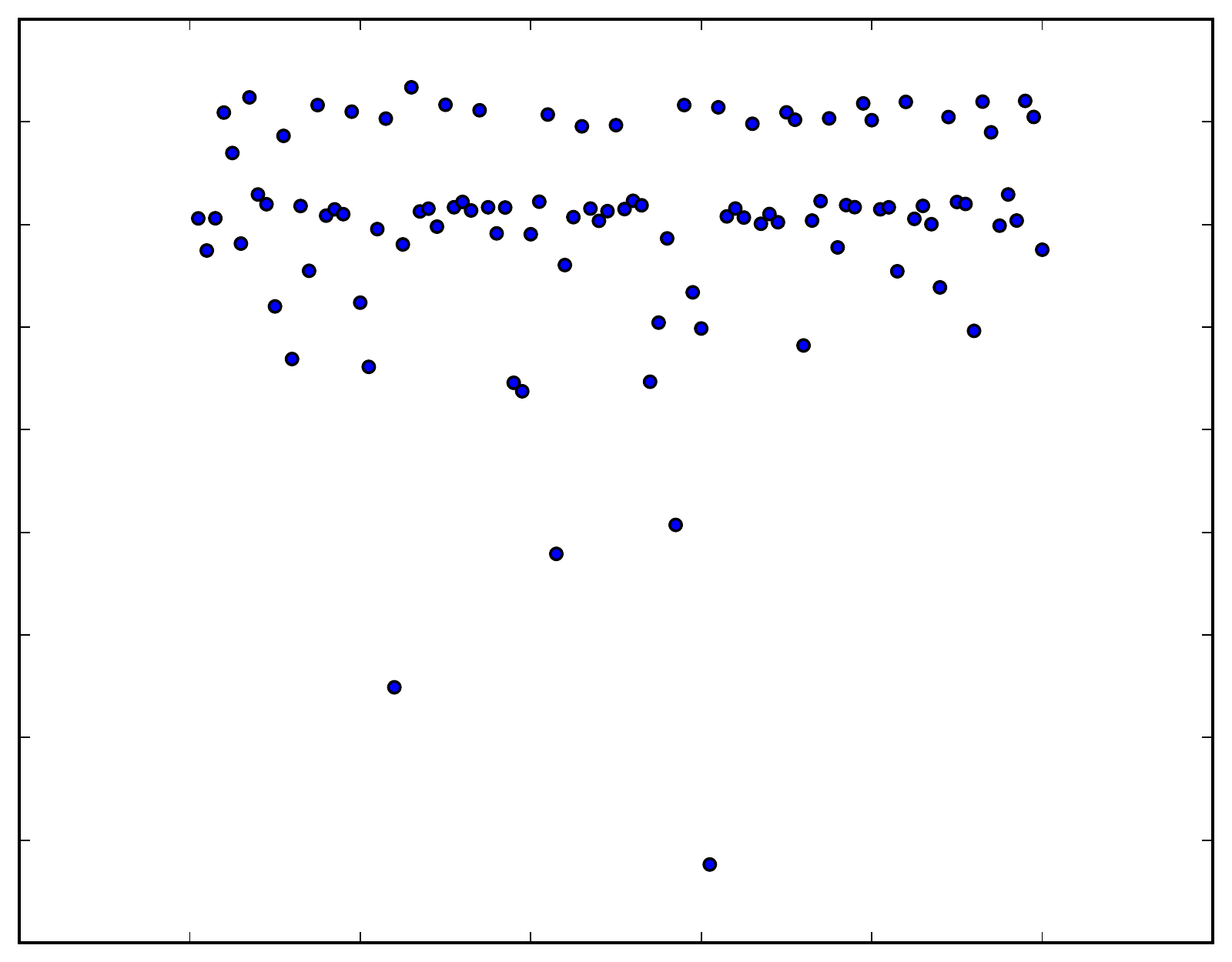}
    \caption{Benchmark 1: Unimodal with skew and large outliers}
\end{subfigure}%
~
\begin{subfigure}{0.22\textwidth}
    \centering
    \includegraphics[width=\textwidth]{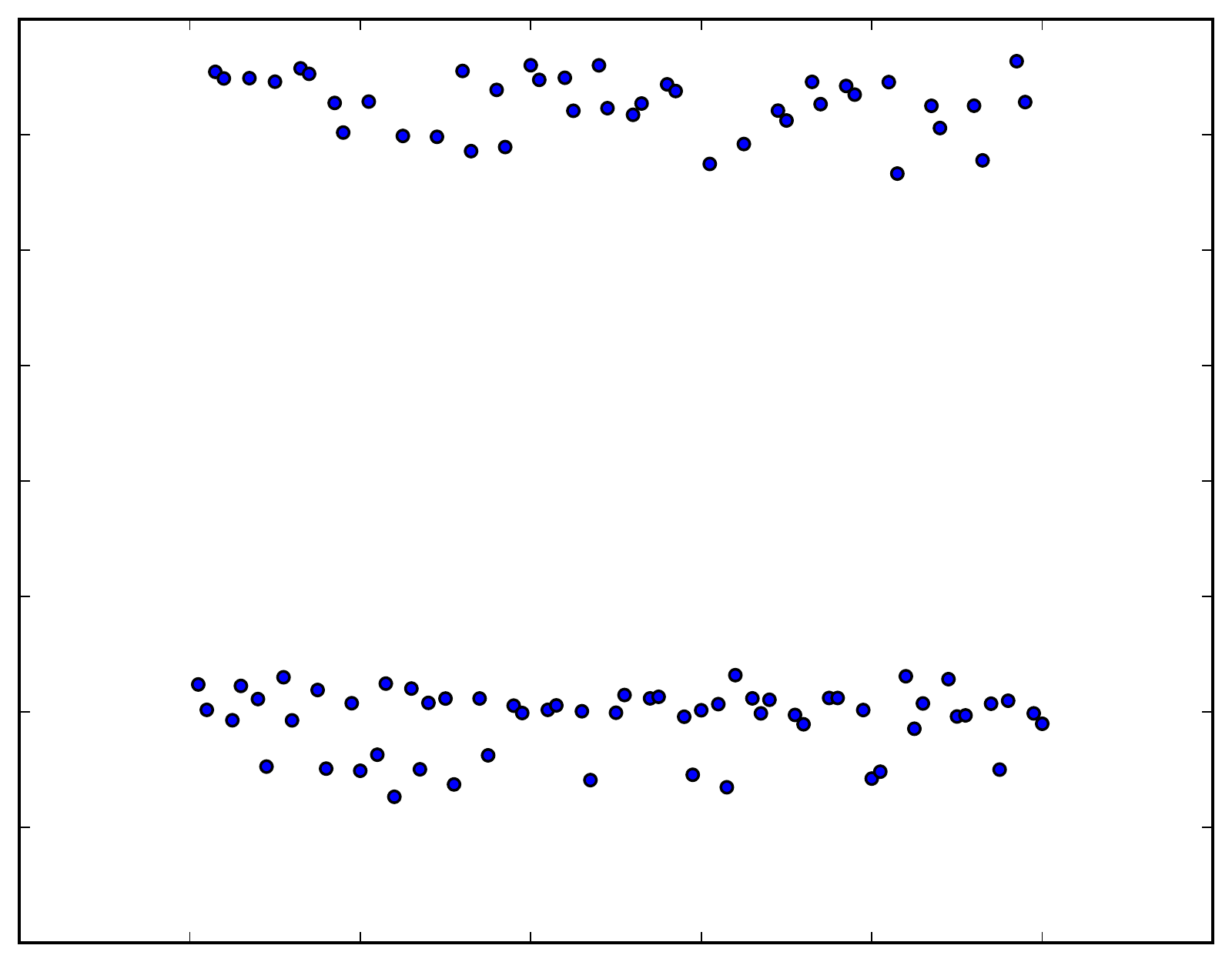}
    \caption{Benchmark 2: Bimodal}
\end{subfigure}
\begin{subfigure}{0.22\textwidth}
    \centering
    \includegraphics[width=\textwidth]{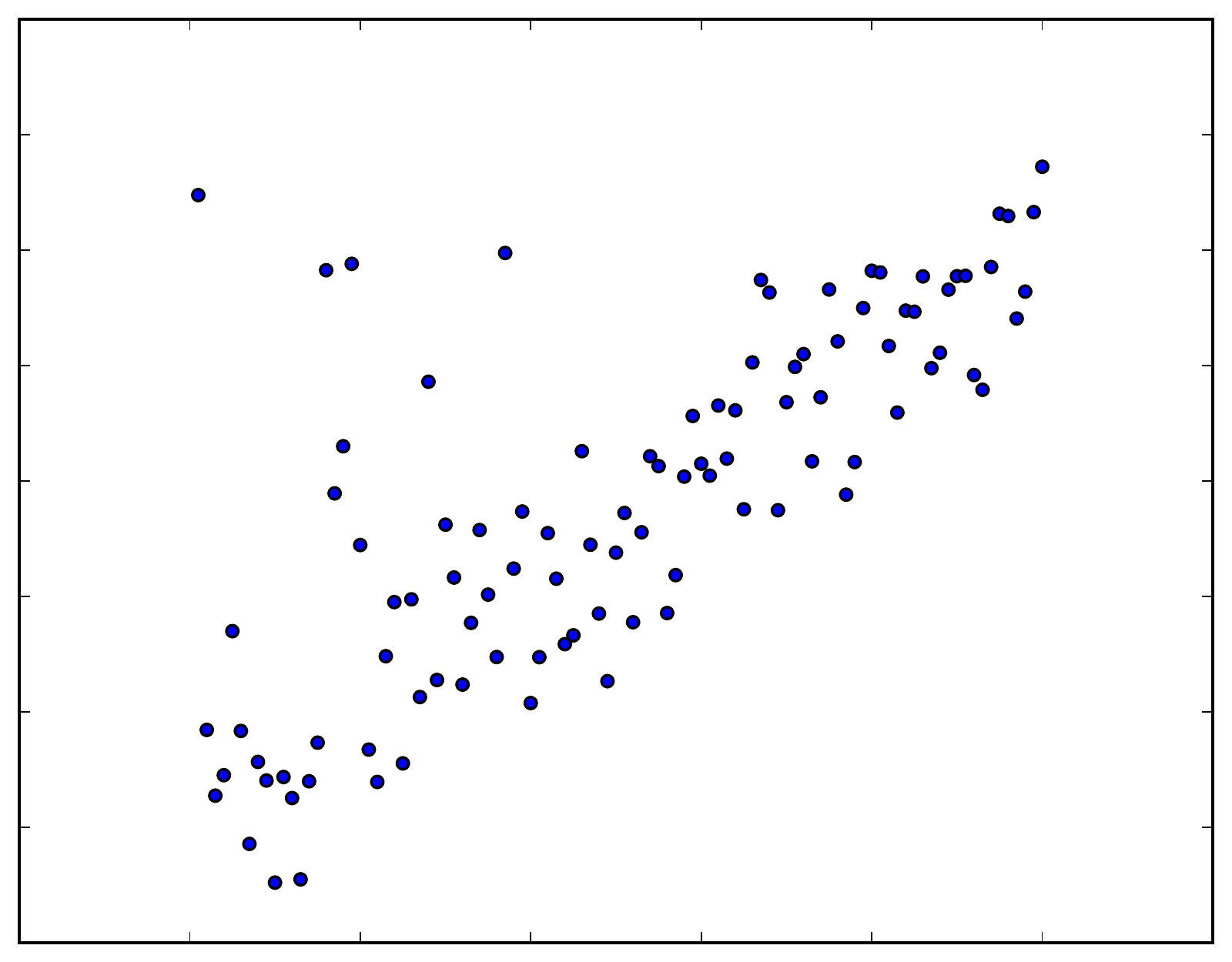}
    \caption{Benchmark 3: Drift}
\end{subfigure}
~
\begin{subfigure}{0.22\textwidth}
    \centering
    \includegraphics[width=\textwidth]{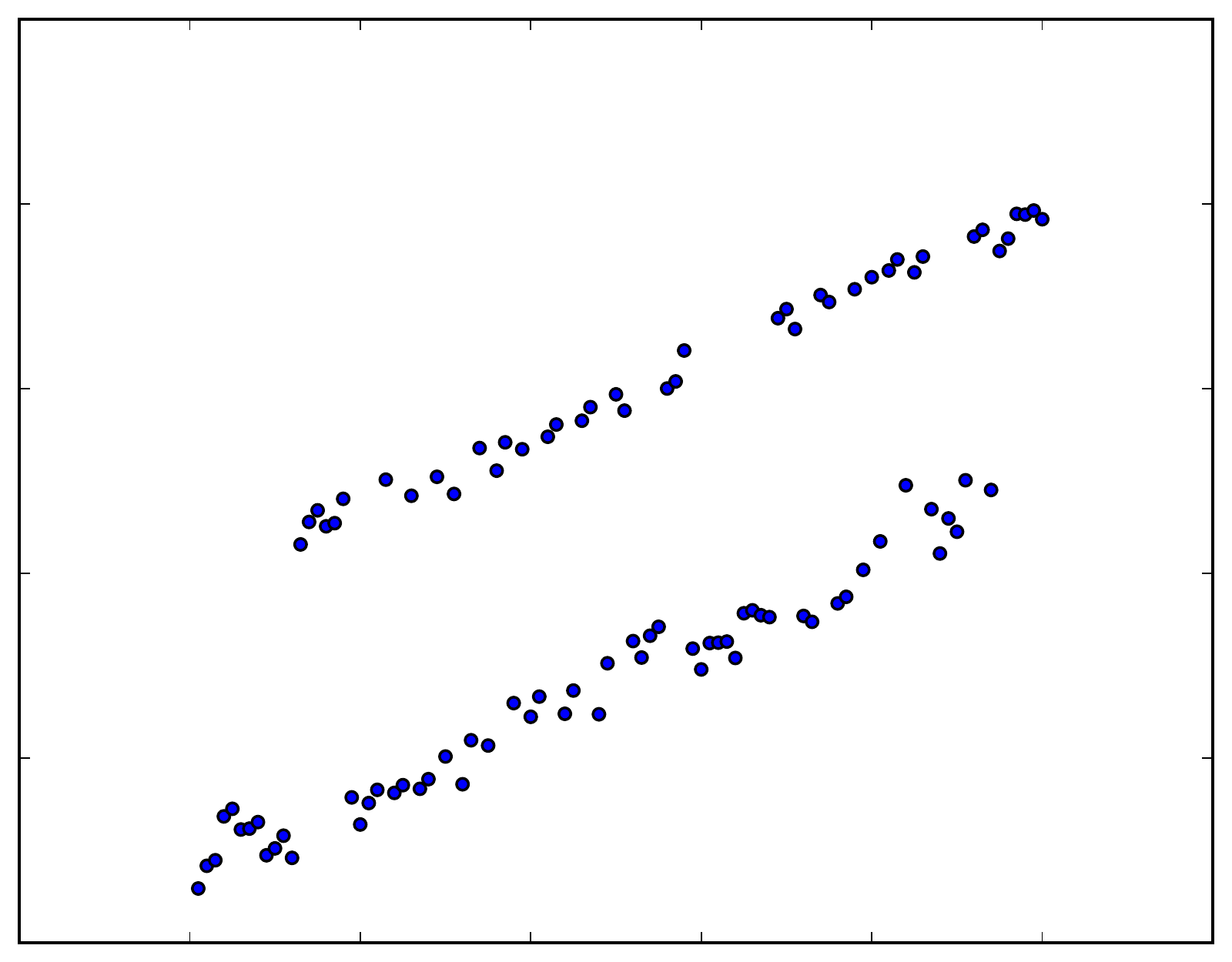}
    \caption{Benchmark 4: Bimodal with drift}
\end{subfigure}
\caption{Variability in the mean benchmark time across multiple trials, showing
that the mean has non-i.i.d., non-normal behavior in four different benchmarks.
Each point represents a mean time computed from trial of 10,000 measurements.
The horizontal axis is the index of the trial, while the vertical axis is time.}
\label{fig:meandistributions}
\vspace{-0.45cm} 
\end{figure}

The existence of many sources of performance variation result in timing
measurements that are not necessarily independent and identically distributed
(i.i.d.). As a result, many textbook statistical approaches fail due to
reliance on the central limit theorem, which does not generally hold in the
non-i.i.d. regime. In particular, empirical program timing distributions are
also often heavy-tailed, and hence contain many outliers that distort measures
of central tendency like the mean, and are not captured in others like the
median.

The violation of the central limit theorem can be seen empirically in many Julia benchmarks.
For example, Figure~\ref{fig:meandistributions} shows that none of the four illustrative
benchmarks considered in this paper exhibit normality in the sample mean. Instead, we see
that the mean demonstrates skewed density and outliers in the first benchmark, bimodality in
the second and fourth benchmarks, and upward drift in the third and fourth benchmarks.

Many other authors have also noted the lack of textbook statistical behavior in timing
measurements~\cite{Gil2011,Chen2015,Rehn2015,Barrett2016}. Authors have also noted the poor
stastistical power of standard techniques such as $F$-tests or Student t-tests for benchmark
timings~\cite{Lilja2000,Mytkowicz2009,Kalibera2013,Chen2015,Barrett2016}. Parametric outlier
detection techniques, such as the 3-sigma rule used in benchmarking software like
\lstinline|AndroBench|\cite{Kim2012}, can also fail when applied to non-i.i.d. timing measurements.

There is a lack of consensus over how non-ideal timing measurements should be treated. Some
authors propose automated outlier removal and analyzing the remaining bulk
distribution~\cite{Kim2012}; however, these methods run the risk of fundamentally distorting
the true empirical distribution for the sake of normal analysis. Other authors have proposed
purposely introducing randomness in the form of custom OS kernels
\cite{Tessellation,Akkan2012}, custom compilers providing reproducible~\cite{Georges2008} or
consistently randomized~\cite{Curtsinger2013} binary layouts, or low-variability garbage
collectors~\cite{Huang2004}. Unfortunately, these methods are specific to a single
programming language, implementation, and/or platform. Furthermore, these methods often
require administrative privileges and drastic modifications to the benchmarking environment,
which are impractical to demand from ordinary users.

\subsection{Existing benchmarking methodologies}
\label{sec:existingtools}

While it is impossible to eliminate performance variation
entirely~\cite{Alcocer2015,Barrett2016}, benchmarking methodologies that attempt to account
for both measurement error and external sources of variation do exist. For example, the
Haskell microbenchmarking package \lstinline|criterion|~\cite{criterion} attempts to thwart
error due to timer inaccuracy by growing the number of benchmark executions per timing
measurement as more timing measurements are obtained. After all measurements are taken, a
summary estimate of the benchmark run time is obtained by examining the derivative of the
ordinary least squares regression line at the point of a single evaluation point. There are
three disadvantages to this approach. First, the least squares fit is sensitive to
outliers~\cite{Maronna2006} (though \lstinline|criterion| does warn the user if outliers are
detected). Second, measurements made earlier in this experiment are highly vulnerable to
timer error, since few benchmark repetitions are used. These early measurements can skew the
regression, and hence also skew the final run time estimate. Third, measurements made later in
the experiment can repeat the benchmark more times than are necessary to overcome timer
error, constituting an inefficient use of experiment time.

Another approach focuses on eliminating ``warm-up'', assuming that first few
runs of a benchmark are dominated by transient background events that
eventually vanish and the timing measurements eventually become
i.i.d.~\cite{Kalibera2013}. Their approach is largely platform-agnostic,
recognizes the pitfalls of inter-measurement correlations, and
acknowledges that merely increasing the number of benchmark repetitions is not always a
sufficient strategy to yield i.i.d. samples. However, the assumption (based on
common folklore) that benchmarks exhibit warm-up is often false, as is clear
from Fig.~\ref{fig:meandistributions} and elsewhere~\cite{Barrett2016}:
even Ref.~\cite{Kalibera2013} itself resorts to ad hoc judgment to work
around the lack of a distinct warm-up phase. There is also no reason to believe that even if warm-up were observed, that the post-warm-up timings will be i.i.d.
Furthermore, the authors do not report if their statistical tools generate
correct confidence intervals. The moment-corrected formulae described are
accurate only for near-normal distributions, which is unlikely to hold for the
kinds of distributions we observed in real world statistics. Additionally, the
methodology requires a manual calibration experiment to be run for each
benchmark, compiler, and platform combination. As a result, this method is is
difficult to automate on the scale of Julia's standard library benchmark suite,
which contains over 1300 benchmarks, and is frequently expanded to improve
performance test coverage.

Below, we describe our methodology to benchmarking for detecting performance regressions,
and how it is justified from a microscopic model for variations in timing measurements. To
the best of our knowledge, our work is the first benchmarking methodology that can be fully
automated, is robust in its assumption of non-i.i.d. timing measurement statistics, and
makes efficient use of a limited time budget.

\section{Terms and definitions}
\label{sec:notation}

\begin{itemize}
    \item
    $P_0$, $P$, $Q_0$, and $Q$ denote \textbf{benchmarkable programs}, each
    defined by a tape (sequence) of instructions.

    \item
    $I^{[i]}_{P}$ is the $i^{\textrm{th}}$ \textbf{instruction} in the tape
    defining program $P$.
    Instructions are indexed in bracketed superscripts, $\cdot^{[i]}$.

    \item
    $D^{[i]}_{P}$ is the \textbf{delay instruction} associated with $I^{[i]}_{P}$.
    Delay instructions are defined in Sec.~\ref{sec:model}.

    \item
    $T_i$ is a \textbf{timing measurement}, namely the amount of time taken to
    perform $n_i$ \textbf{executions} of a benchmarkable program. This quantity
    is directly measurable in an experiment.

    \item
    $t$ is a \textbf{theoretical execution time}.
    $t_{P_0}$ is the minimum time required to perform a single execution of
    $P_0$ on a given computer.

    \item
    \textbf{Estimated quantities} are denoted with a hat, $\hat\cdot$.
    For example, $\hat{t}_{P_0}$ is an estimate of the theoretical execution
    time $t_{P_0}$.

    \item
    A benchmark \textbf{experiment} is a recipe for obtaining multiple timing
    measurements for a benchmarkable program. Experiments can be executed to
obtain \textbf{trials}. The
    $i^{\textrm{th}}$ trial of an experiment is a collection of timing measurements
    $T^{\{i\}}_1, \dots T^{\{i\}}_j, \dots T^{\{i\}}_k$. Trial indices are always
    written using embraced superscripts, $\cdot^{\{i\}}$.

    \item
    $\tau$ denotes time quantities that are external to the benchmarkable program:
    \begin{itemize}
        \item $\tau_{\textrm{budget}}$ is the \textbf{time budget} for an experiment.
        \item $\tau_{\textrm{acc}}$ is the \textbf{accuracy} of the system timer, i.e.\ an upper bound on the maximal error in using the system timer to time an experiment.
        \item $\tau_{\textrm{prec}}$ is the \textbf{precision} of the system timer, namely the smallest nonzero time interval measurable by the timer.
    \end{itemize}

    \item
    $x_P^{(i)[j]} \tau^{(i)}$ is the \textbf{time delay} due to the
    $i^{\textrm{th}}$ \textbf{delay factor} for delay instruction $D^{[j]}$.  Specifically,
    $\tau^{(i)}$ is the factor's \textbf{time scale} and $x_P^{(i)[j]}$ is the factor's
    \textbf{trigger coefficient}, as introduced Sec.~\ref{sec:model}. Delay factors are indexed with  parenthesized
    superscripts, $\cdot^{(i)}$.

    \item
    $\epsilon$ is the measurement error due to timer inaccuracy.

    \item
    $E_m = \frac{T_m}{n_m} - t_{P_0}$ is the total contribution of all delay
factors found in measurement $m$, plus the measurement error $\epsilon$.

    \item
    $X^{(i)}_P$ is the \textbf{total trigger count} of the $i^{\textrm{th}}$
    delay factor during the execution of program $P$.

    \item
    $\nu$ is an \textbf{oracle function} that, when evaluated at an execution
time $t$, estimates an appropriate $n$ necessary to overcome measurement error
due to
    $\tau_{\textrm{acc}}$ and $\tau_{\textrm{prec}}$. The oracle function is described in detail in Sec.~\ref{sec:oracle}.
\end{itemize}

\section{A model for benchmark timing distributions}
\label{sec:model}

We now present a statistical description of how benchmark programs behave
when they are run in serial. Our model deliberately avoids the problematic
assumption that timing measurments are i.i.d. We will use this model later to
justify the design of a new automated experimental procedure.

\subsection{User benchmarks run with uncontrollable delays}
\label{sec:programmodel}

Let $P_0$ be a deterministic benchmark program which consists of an instruction tape
consisting of $k$ instructions:
\begin{equation}
    P_0 = \left[I^{[1]}, I^{[2]}, \dots I^{[k]}\right].
\end{equation}
Let $\tau^{[i]}$ be the run time of instruction $I^{[i]}$. Then, the total run
time of $P_0$ can be written $t_{P_0} = \sum_{i=1}^N \tau^{[i]}$.

While a computer may be directed to execute $P_0$, it may not necessarily run the program's
instructions as they are originally provided, since the environment in which $P_0$
runs is vulnerable to the factors described earlier in Sec.~\ref{sec:variations}. Crucially,
these factors only \textit{delay} the completion of the original instructions, rather than
speed them up.\footnote{While there are a very few external factors which might speed up
program execution, such as frequency scaling~\cite{RHEL6}, they can be easily accounted for
by ensuring that power consumption profiles are always set for maximal performance. We
therefore assume that these factors have been accounted for.} Therefore, we call them
\textit{delay factors}; they can be modeled as extra instructions which, when
interleaved with the original instructions, do not change the semantics of
$P_0$, but still add to the program's total run time. Thus, we can define a new
program $P$ which consists of $P_0$'s original instructions interleaved with
additional \textit{delay instructions} $D^{[i]}$:
\begin{equation}
    P = \left[I^{[1]}, D^{[1]}, I^{[2]}, D^{[2]}, \dots I^{[k]}, D^{[k]}\right].
\end{equation}
The run time of $P$ can then be written
\begin{equation}
    t_P = t_{P_0} + \sum_{i} \tau^{[i]}_D,
\end{equation}
where $\tau^{[i]}_D$ is the execution time of $D^{[i]}$. Since $\tau^{[i]}_D \ge 0$, it
follows that $t_P \ge t_{P_0}$.

The run time of each delay instruction, $\tau^{[i]}_D$, can be further decomposed
into the runtime contributions of individual delay factors. Let us imagine that
each delay factor $j$ can either contribute or not contribute to $D^{[i]}$.
Assuming that each delay factor triggers inside $D^{[i]}$ with constant
probability $p^{[i](j)}$ of taking a fixed time $\tau^{(j)}$, we can then
write:
\begin{equation}
    \tau^{[i]}_D = \sum_{i} x_P^{[i](j)} \tau^{(j)},
\end{equation}
where $x_P^{[i](j)}$ is a Bernoulli random variable with success probability $p^{[i](j)}$.
We denote the total number of times the $i^{\textrm{th}}$ delay factor was triggered during
the execution of $P$ as the \textit{trigger count} $X_P^{(i)} = \sum_{j} x_P^{[i](j)}$.
Since the trigger count is a sum of independent Bernoulli random variables with nonidentical
success probabilities, $X_P^{(j)}$ is itself a random variable that follows a Poisson
binomial distribution parameterized by the success probabilities $\left[p^{(1)[j]}, \dots
p^{(k)[j]}\right]$. Our final expression for $t_P$ in terms of these quantities is then:
\begin{align}
t_P &= t_{P_0} + \sum_{i=1}^{k} \tau^{[i]}_D \nonumber \\
    &= t_{P_0} + \sum_{i=1}^{k} \sum_{j} x_P^{[i](j)} \tau^{(j)} \nonumber \\
    &= t_{P_0} + \sum_{j} X_P^{(j)} \tau^{(j)}.
\end{align}
In summary, our model treats $t_P$ as a random variable whose distribution
depends on the trigger probabilities $p^{[i](j)}$, which are determined by the
combined behavior of the delay factors and the initial benchmark program $P_0$.

\subsection{Repeated benchmark execution is often necessary but not always sufficient}
\label{sec:measuremodel}

As mentioned in Sec.~\ref{sec:existingtools}, experiments which measure program
performance usually incorporate multiple benchmark executions to obtain more
accurate measurements. We now apply our model to show that multiple executions
are necessary to eliminate error due to timer inaccuracy, but are insufficient
to obviate delay factors.

Represent $n$ executions of the program $P_0$ comprised of $k$ instructions as
a single execution of a program $Q_0$, which is the result of concatenating $n$
copies of $P_0$:
\begin{align}
Q_0 &= \left[P_0, P_0, \dots P_0 \right] \nonumber \\
    &= \left[I_{P}^{[1]}, \dots I_{P}^{[k]}, I_{P}^{[1]}, \dots I_{P}^{[k]}, I_{P}^{[1]}, \dots I_{P}^{[k]} \right] \nonumber \\
    &= \left[I_{Q}^{[1]}, I_{Q}^{[2]}, \dots I_{Q}^{[nk]} \right],
\end{align}
with $I_{P}^{[i]} = I_{Q}^{[i + ck]}$ for $c \in \{0, \dots n - 1\}$. The
subscripts on $I$ denote the program which contains that instruction, with the
0 subsubscript dropped for brevity.

Now interleave delay instructions as before to obtain the program $Q$ that is
actually executed. $Q$ is \textit{not} simply $n$ repetitions of $P$, since the
delay instructions in
$Q$ are not simply copies of the delay instructions in $P$. An observed timing measurement
$T$ of a single execution of $Q$ can be decomposed as:
\begin{align}
    T &= t_{Q} + \epsilon \nonumber \\
      &= t_{Q_0} + \sum_{j} X_Q^{(j)} \tau^{(j)} + \epsilon \nonumber \\
      &= n \, t_{P_0} + \sum_{j} \sum_{i=1}^{nk} x_Q^{[i](j)} \tau^{(j)} + \epsilon,
\end{align}
where $\epsilon$ is the error due to timer inaccuracy (whose magnitude must by definition be smaller than $\tau_\textrm{acc}$).

We may try to determine $t_{P_0}$ from the experimental time as $T/n$, which is
also the gradient of a linear model for $T$ against $n$ when the intercept is
zero. However, our model gives instead:
\vspace{-0.10cm}
\begin{equation}
    \frac{T}{n} = t_{P_0} + \frac{\sum_{j} \sum_{i=1}^{nk} x_Q^{[i](j)} \tau^{(j)} + \epsilon}{n}.
\end{equation}
All the terms on the right hand side other than $t_{P_0}$ constitute the
error $E$ in our measurement. For large $n$, the term $\epsilon/n$ arising from
timer inaccuracy becomes negligible, but the behavior of the other term depends
on the specific structure of the delay factors. In the best case, each delay
factor triggers $o(n)$ times, so that $T/n\to t_{P_0}$ as desired. However, in
the worst case, every factor triggers on every instruction, $x_Q^{[i](j)}
= 1$, and the large $n$ behavior of $T/n$ does not reduce to the true run time
$t_{P_0}$, but rather:
\begin{equation}
\label{eq:11}
    \lim_{n\to\infty} \frac{T}{n} = t_{P_0} + k \sum_{j} \tau^{(j)}.
\end{equation}

\eqref{eq:11} is a key result of our model: one cannot always reliably
eliminate bias due to external variations simply by executing the benchmark
many times. Whether or not increasing $n$ can render the delay factor term
negligible depends entirely on the distribution of trigger counts $X_Q^{(j)}$,
which are difficult or impossible to control (see Sec.~\ref{sec:variations}).
Therefore, we can only expect that $T/n$ at large $n$ gives us at best an
\textit{overestimate} of the true run time $t_{P_0}$.

\section{An automated procedure for configuring performance experiments}
\label{sec:confexperiment}

In this section, we present an experimental procedure for automatically
selecting useful values of $n$ for a given benchmark program, which can be
justified from our model of serial benchmark execution above. Our procedure
estimates a value for $n$ which primarily minimizes error in timing
measurements and secondarily maximizes the number of measurements obtainable
within a given time budget.

\subsection{An algorithm for estimating the optimal $n$ value}

Given $P_0$ and a total time budget $\tau_{\textrm{budget}}$, we use the automatable
procedure in Alg.~\ref{alg:tuning} for guessing the minimum value of $n$ required to
amortize measurement error due to timer inaccuracy. The algorithm makes use of
an oracle function $\nu$, which is discussed in greater detail below in
Sec.~\ref{sec:oracle}.

\begin{algorithm}
    \caption{Estimating $n$, the optimal number of benchmark repetitions required to
    minimize timer error and maximize the number of data points obtainable within a
    time budget.}
    \label{alg:tuning}
    \KwIn{$P_0$, $\tau_{\textrm{acc}}$, $\tau_{\textrm{prec}}$, an oracle function $\nu : t \to n$}
    \KwOut{$n$}
    Let $j = \tau_{\textrm{acc}} / \tau_{\textrm{prec}}$.

    For $i \in \{1, \dots j\}$, measure the amount of time it takes to perform $i$
    executions of $P_0$, resulting in a collection of timing measurements $T_1, \dots T_j$.

    Estimate $t_{P_0}$ as $\hat{t}_{P_0} = \textrm{min}(\frac{T_1}{1}, \dots \frac{T_j}{j})$.

    Evaluate $\nu(\hat{t}_{P_0})$ to obtain $n$. Details of $\nu$ are given in Sec.~\ref{sec:oracle}
\end{algorithm}

The upper bound $j$ in Alg.~\ref{alg:tuning} is the ratio of timer accuracy to
timer precision. If each timing measurement consists of more than $j$
repetitions, then the contribution of timer inaccuracy to the total error is
less than $\tau_\textrm{acc} / j = \tau_\textrm{prec}$, and so is too small to
measure. Thus, there is no reason to pick $n > j$. In practice,
$\tau_\textrm{acc}$ need only be an overestimate for the timer accuracy, which
would raise the $n$ determined, but is still an acceptable result.

Alg.~\ref{alg:tuning} need only be applied once per benchmark, since the
estimated $n$ can be cached for use in subsequent experiments on the same
machine. Thus, we consider this algorithm an automated preprocessing step that
does not count against our time budget $\tau_{\textrm{budget}}$. In this
regard, our approach differs significantly from other approaches like
\lstinline|criterion|, which re-determines $n$ every time a benchmark is run.

\subsection{Justifying the minimum estimator}
\label{sec:minimum}

\begin{figure}
\centering
\includegraphics[width=\columnwidth]{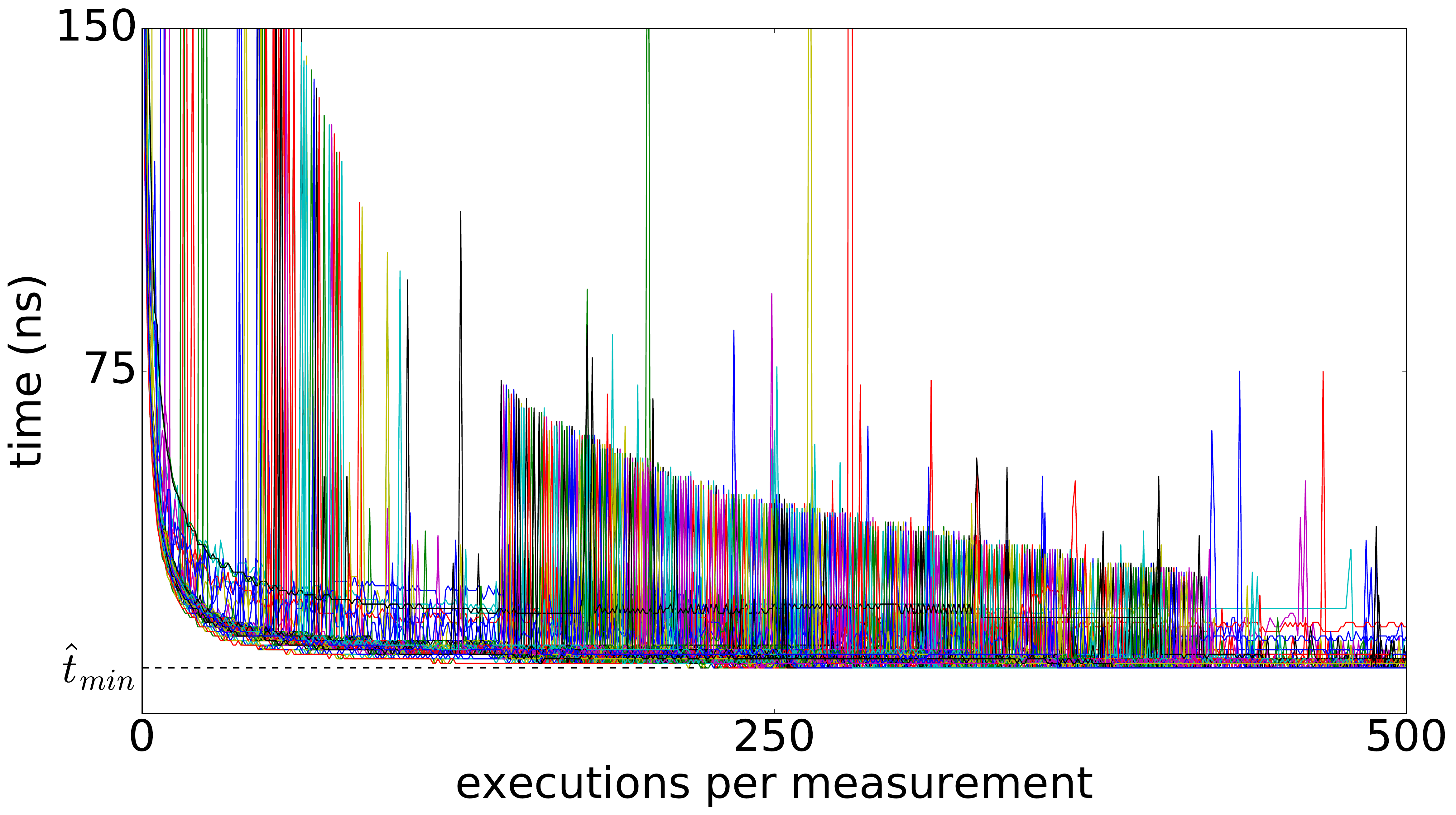}
\caption{Plots of $T/n$ vs.\ $n$ produced by repeated experiments, each
consisting of running Alg.~\ref{alg:tuning} on the \lstinline|branchsum|
benchmark. While each experiment can produce wildly oscillatory curves, the
minimum across all the curves at each $n$ is much smoother and asymptotically
tends toward the same constant value.}
\label{fig:scaling}
\end{figure}

We will now justify Alg.~\ref{alg:tuning}'s use of the minimum to estimate $t_{P_0}$, as
opposed to the more common median or mean.
Consider the total error term for a given timing measurement $E_m = \left(\sum_{i}
X_Q^{(i)} \tau^{(i)} + \epsilon \right)_m / n_m$, such that $T_i/n_i = t_{P_0} +
E_i$. The minimum estimator applied to our timing measurements can then be
written as:

\begin{align}
    \hat{t}_{P_0} &= \textrm{min}(\frac{T_1}{1}, \dots \frac{T_j}{j}) \nonumber \\
                  &= t_{P_0} + \textrm{min}(E_1, \dots E_j).
\end{align}
Thus, $\hat{t}_{P_0}$ is the estimate of $t_{P_0}$ which minimizes the error terms appearing
in our sample.

In the limit where the delay factor time scales are greater than $\tau_{\textrm{acc}}$, the
total error terms will always be positive, such that choosing the smallest timing
measurement will choose the sample with the smallest magnitude of error. If the delay factor
time scales are less than $\tau_{\textrm{acc}}$, choosing the smallest timing measurement
might choose a sample which underestimates $t_{P_0}$ due to negative timer error. In this
case, Alg.~\ref{alg:tuning} will simply pick a larger $n$ than is strictly necessary, which
is still acceptable.

\begin{figure}
\centering
\includegraphics[width=\columnwidth]{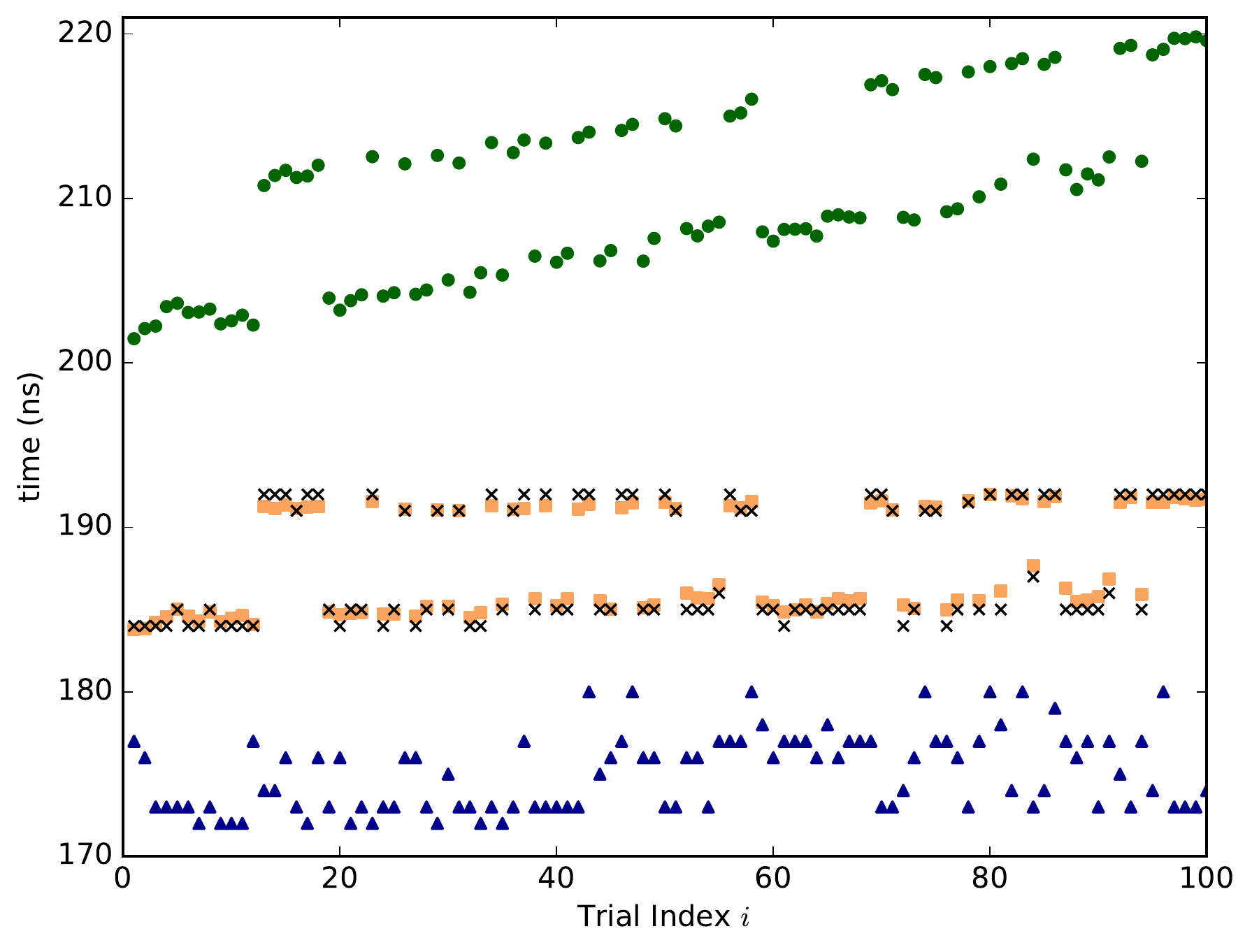}
\caption{The behavior of different location parameters across multiple trials of
the \lstinline|sumindex| benchmark: mean (green filled circles), trimmed mean of
the 5th---95th percentiles (brown filled squares), median (black crosses), and
minimum (blue filled triangles).}
\label{fig:locationmeasures}
\end{figure}

\begin{figure}
\centering
\includegraphics[width=\columnwidth]{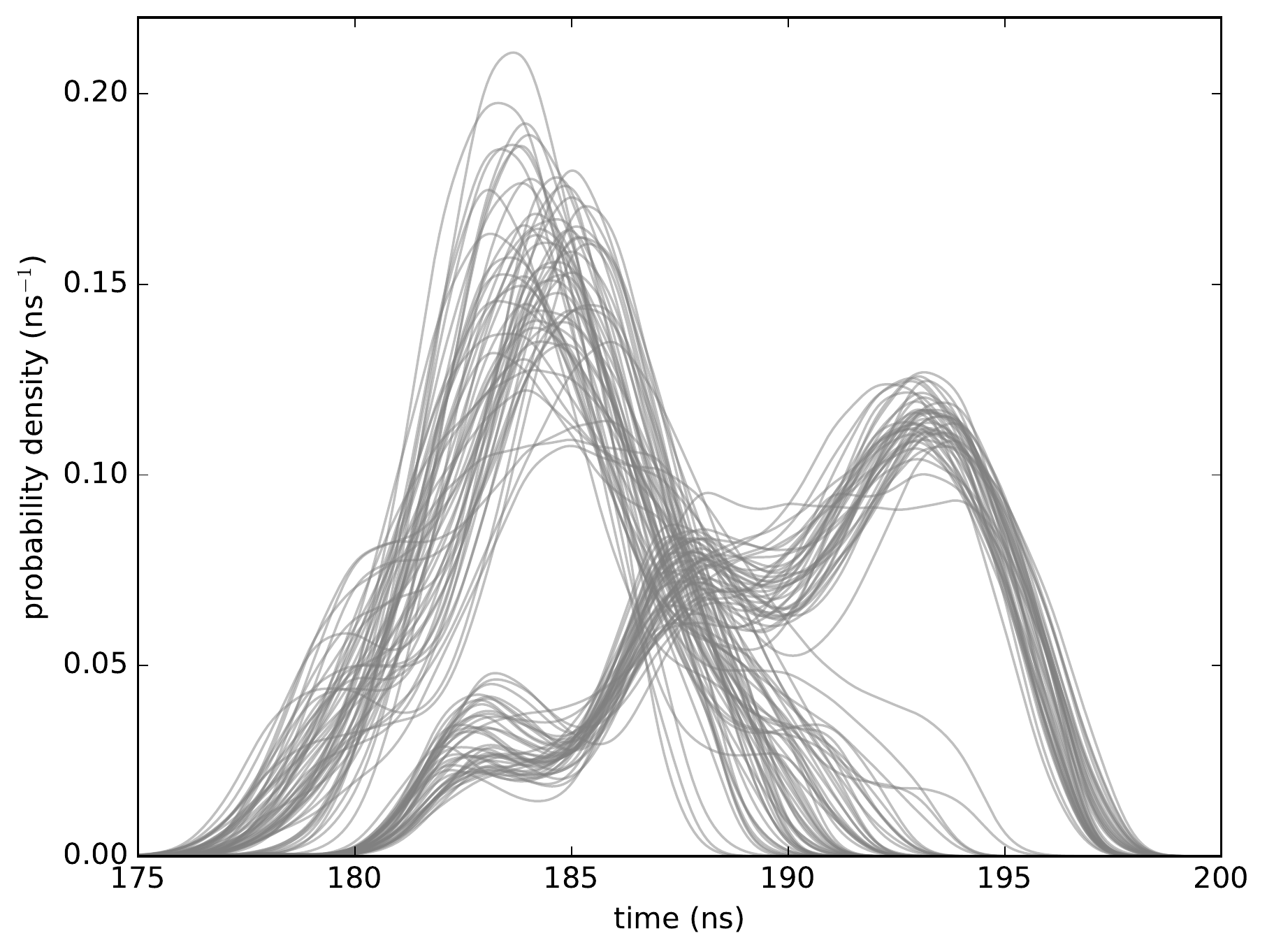}
\caption{Kernel density estimates (KDEs) of the probability density functions
(pdfs) across 100 trials of the \lstinline|sumindex| benchmark. Each curve is a
KDE formed from a trial of 10,000 consecutively gathered timing measurements.
Note that the data form two distinct clusters. A cursory investigation did not
reveal any inter-trial correlations that revealed a predictable preference for
which cluster would be observed.}
\label{fig:pdfsumindex}
\end{figure}

Figs.~\ref{fig:locationmeasures} and \ref{fig:pdfsumindex} provide further
justification for the minimum over other common estimators like the median,
mean, or trimmed mean. Recall from Section~\ref{sec:model} that the error terms
$E_i$ are sampled from a sum of scaled random variables following nonidentical
Poisson binomial distributions. As such, these terms can and do exhibit
multimodal behavior. While estimators like the median and trimmed mean are
known to be robust to outliers~\cite{Maronna2006},
Fig.~\ref{fig:locationmeasures} demonstrates that they still capture bimodality
of the distributions plotted in Fig.~\ref{fig:pdfsumindex}. Thus, these
estimators are undesirable for choosing $n$, since the result could vary
drastically between different executions of Alg.~\ref{alg:tuning}, depending on
which of the estimator's modes was captured in the sample, and hence affect
reproducibility. In contrast, the distribution of the minimum across all
experimental trials is unimodal in all cases we have observed. Thus for our
purposes, the minimum is a unimodal, robust estimator for the location
parameter of a given benchmark's timing distribution.

\subsection{The oracle function}
\label{sec:oracle}

Our heuristic takes as input an oracle function $\nu(t)$ that maps expected
run times to an optimal number of executions per measurement. While
Alg.~\ref{alg:tuning} does not directly describe $\nu(t)$, appropriate choices
for this function should have the following properties:

\begin{itemize}
    \item $\nu(t)$ has a discrete range $\{1, \dots, j\}$.
    \item $\nu(t)$ is monotonically decreasing, so that the longer the run time, the fewer repetitions per measurement.
    \item $\frac{d\nu}{dt}|_{t \approx \tau_{\textrm{prec}}} \approx 0$,
    so that there is only weak dependence on the timer precision parameter,
    which may not be accurately known.
    \item $\frac{d\nu}{dt}|_{t \approx \tau_{\textrm{acc}}} \approx 0$,
    so that there is only weak dependence on the timer accuracy parameter,
    which may not be accurately known.
    \item $\nu(\tau_{\textrm{prec}}) \approx j$, so that benchmarks that take a     short time to run are not repeated more times than necessary to mitigate timer inaccuracy.
    \item $\nu(t \ge \tau_{\textrm{acc}}) \approx 1$, so that benchmarks that take a long time to run need not be repeated.
\end{itemize}

There are many functions that satisfy these criteria. One useful example takes
the form of the generalized logistic function:

\begin{equation}
\label{eq:glog}
    Y(t) = \floor*{1 + \frac{j - 1}{1 + e^{a (t - b \tau_{\textrm{acc}})}}}
\end{equation}
where reasonable values of $a$ and $b$ are approximately $0.005 < a \tau_{\textrm{prec}} < 0.02$ and $0.4 < b <
0.6$.

\begin{figure}
\centering
\includegraphics[width=\columnwidth]{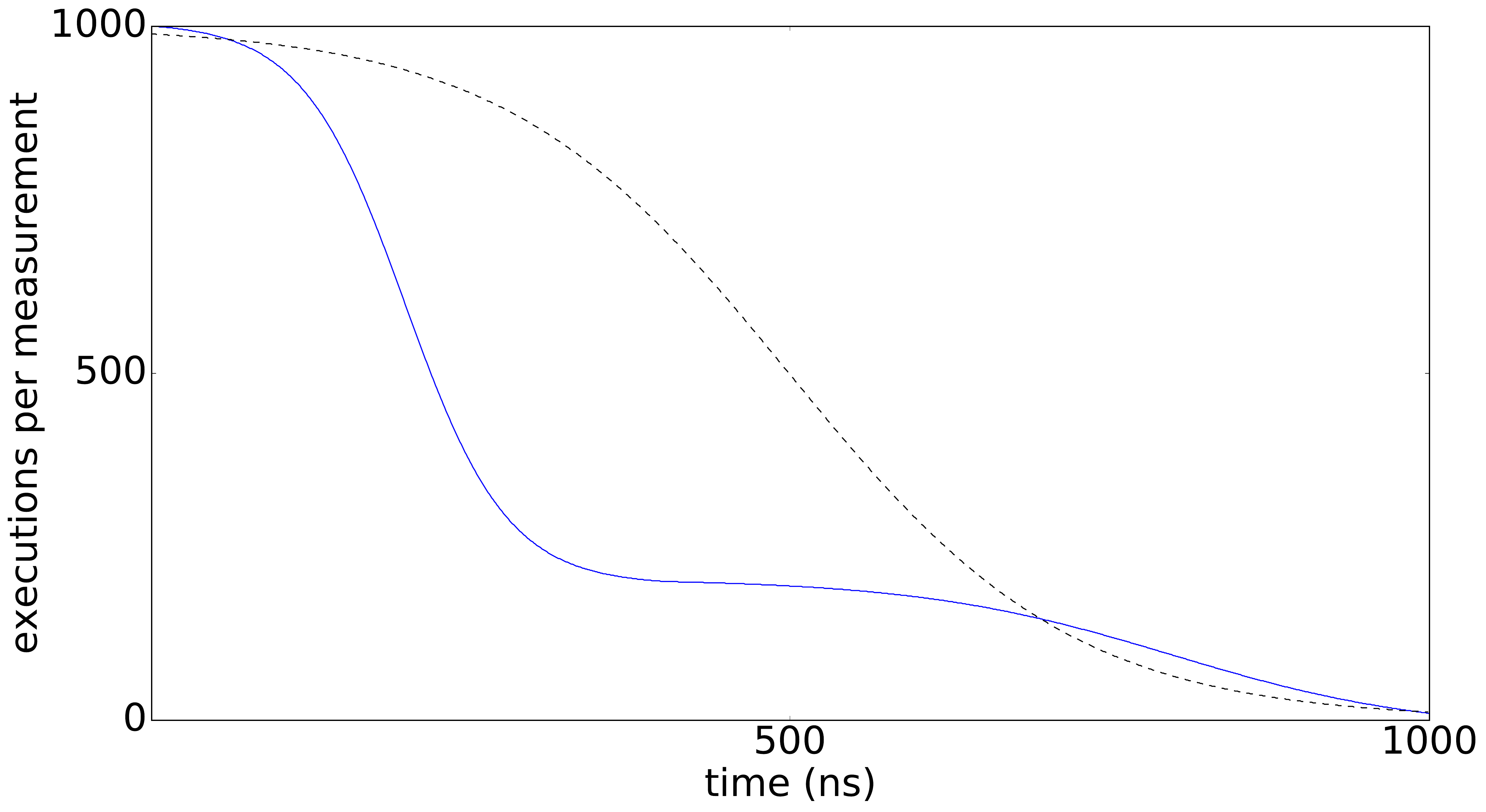}
\caption{Two possible oracle functions for $\nu(t)$ at $\tau_{\textrm{acc}} \approx 1000 \textrm{ns}$, $\tau_\textrm{prec} \approx 1 \textrm{ns}$.
The solid blue curve is an example of an empirically tuned lookup table, while
the dotted black curve is $Y(t)$ from Eq~\ref{eq:glog} with parameters $a =
0.009 / \tau_\textrm{prec}$ and $b = 0.5$.}
\label{fig:oracle}
\end{figure}

In practice, we have found that better results can be achieved by first
approximating $Y(t)$ with a lookup table, then modifying the lookup table based
on empirical observations. This was accomplished by examining many benchmarks
with a variety of known run times at different time scales, seeking for each
run time the smallest $n$ value at which the minimum estimate appears to
converge to a lower bound (e.g.\ around $n = 250$ for the benchmark in
Fig.~\ref{fig:scaling}). Fig.~\ref{fig:oracle} plots both \eqref{eq:glog} and an
empirically obtained lookup table as potential oracle functions.

\section{Implementation in Julia}
\label{sec:implementation}

The experimental methodology in this paper is implemented in the
\lstinline|BenchmarkTools| Julia
package\footnote{\url{https://github.com/JuliaCI/BenchmarkTools.jl}}. In
addition to the
\lstinline|BaseBenchmarks|\footnote{\url{https://github.com/JuliaCI/BaseBenchmarks.jl}}
and
\lstinline|Nanosoldier|\footnote{\url{https://github.com/JuliaCI/Nanosoldier.jl}}
packages, the \lstinline|BenchmarkTools| package implements the on-demand CI
benchmarking service used by core Julia developers to compare the performance
of proposed language changes with respect to over 1300 benchmarks. Since this
CI benchmarking service began in early 2016, it has caught and prevented the
introduction of dozens of serious performance regressions into Julia's standard
library (defining a serious regression as a $30\%$ or greater increase in a
benchmark's minimum execution time).

The benchmarks referenced in this paper are Julia benchmarks written and
executed using \lstinline|BenchmarkTools|. A brief description of each
benchmark is offered below:

\begin{itemize}
    \item The \lstinline|sumindex(a, inds)| benchmark sums over all
\lstinline|a[i]| for all \lstinline|i| in \lstinline|inds|. This test stresses
memory layout via element retrieval.
    \item The \lstinline|pushall!(a, b)| benchmark pushes elements from \lstinline|b| into
    \lstinline|a| one by one, additionally generating a random number at each iteration (the
    random number does not affect the output). This test stresses both random number
    generation and periodic reallocation that occurs as part of Julia's dynamic array
    resizing algorithm.
    \item The \lstinline|branchsum(n)| benchmark loops from \lstinline|1| to \lstinline|n|.
    If the loop variable is even, a counter is decremented. Otherwise, an inner loop is
    triggered which runs from \lstinline|1| to \lstinline|n|, in which another parity test
    is performed on the inner loop variable to determine whether to increment or decrement
    the counter. This test stresses periodically costly branching within loop iterations.
    \item The \lstinline|manyallocs(n)| allocates an array of \lstinline|n| elements, where
    each element is itself an array. The inner array length is determined by a random number
    from \lstinline|1| to \lstinline|n|, which is regenerated when each new array is
    constructed. However, the random number generator is reseeded before each generation so
    that the program is deterministic. This test stresses random number generation and
    the frequent allocation of arrays of differing length.
\end{itemize}

The mock benchmark suite referenced in this paper is hosted on GitHub at
\url{https://github.com/jiahao/paper-benchmark}.

\section{Conclusion}
\label{sec:conclusion}

The complexities of modern hardware and software environments produce
variations in benchmark timings, with highly nonideal statistics that
complicate the detection of performance regressions. Timing measurements taken
from real Julia benchmarks confirm the observations of many other authors
showing highly nonideal, even multimodal behavior, exhibited by even the
simplest benchmark codes.

Virtually all timing variations are delays caused by flushing cache lines, task
switching to background OS processes, or similar events. The simple
observation that variations never reduce the run time led us to consider a
straightforward analysis based on a simple model for delays in a serial
instruction pipeline. Our results suggest that using the minimum estimator for
the true run time of a benchmark, rather than the mean or median, is robust to
nonideal statistics and also provides the smallest error.  Our model also
revealed some behaviors that challenge conventional wisdom: simply
running a benchmark for longer, or repeating its execution many times, can
render the effects of external variation negligible, even as the error due to
timer inaccuracy is amortized.

Alg.~\ref{alg:tuning} presents an automatable heuristic for selecting the
minimum number of executions of a benchmark per measurement required to defeat
timer error. This strategy has been implemented in the
\lstinline|BenchmarkTools| Julia package, which is employed daily and on demand
as part Julia's continuous integration (CI) pipeline to evaluate the
performance effects of proposed changes to Julia's standard library in a fully
automatic fashion. \lstinline|BenchmarkTools| can also be used to test the
performance of user-authored Julia packages.

\section*{Acknowledgment}
\label{sec:acknowledgement}

We thank the many Julia developers, in particular Andreas Noack (MIT), Steven
G.\ Johnson (MIT) and John M. White (Facebook), for many insightful
discussions.

This research was supported in part by the U.S. Army Research Office under
contract W911NF-13-D-0001, the Intel Science and Technology Center for Big
Data, and DARPA XDATA.

\bibliography{biblio}
\bibliographystyle{IEEEtran}

\end{document}